\documentclass[sigconf]{acmart}
\AtBeginDocument{%
  }

\setcopyright{acmlicensed}
\copyrightyear{2025}
\acmYear{2025}
\acmConference[ACM SIGSPATIAL '25]{33rd ACM SIGSPATIAL
International Conference on Advances
in Geographic Information Systems}{November 03--06,
  2025}{Minneapolis, MN}

\usepackage{svg}
\usepackage{textcomp}
\usepackage{float}
\usepackage{comment}
\usepackage{algorithm2e}


\begin{document}

\title{A Highly Available GTFS-RT Positions System [System]}

\author{Joshua Wong}
\affiliation{%
  \institution{California State University, Los Angeles}
  \city{Los Angeles}
  \state{California}
  \country{United States}
}
\email{jwong159@calstatela.edu}

\author{Kin Tsang}
\affiliation{%
  \institution{California State University, Los Angeles}
  \city{Los Angeles}
  \state{California}
  \country{United States}
}
\email{ktsang3@calstatela.edu}
\renewcommand{\shortauthors}{Wong et al.}

\begin{abstract}
We develop a system for real-time public transportation data, deciding to use the data standard GTFS-RT (GTFS Realtime), an open data format for public transit data. We give an overview of the design of a physical GPS sensor device, its firmware, and processes. Next, we give the algorithms used to translate raw sensor data into a public GTFS-RT data feed. We deploy this feed over a highly available cluster across multiple regions to maintain high availability.
\end{abstract}




\maketitle

\section{Introduction}
Many small, municipal public transportation agencies do not have a realtime data system. Realtime data for public transit has many positive effects, from increasing ridership \cite{ferris2010onebusaway}, to increasing the satisfaction of the ridership experience \cite{gooze2013benefits}. Unfortunately, there are many factors that contribute to small agencies not having a realtime data system. According to RebelGroup \cite{RebelGroup2024}, one of the primary factors is a lack of resources for transit technology, particularly in the effort required procure contracts. However, one way the State of California has promoted the adoption of certain transit technologies and standards, such as GTFS (General Transit Feed Specification), is through endorsement and support, leading to increased usage and implementation of these endorsed technologies and standards. As a California Public Benefit Nonprofit Corporation, in partnership with a State Government initiative, we will endorse and provide technical assistance for transit
agencies' implementation GTFS real-time data system, leading to increased useage of GTFS-RT (General Transit Feed Specification - Realtime) in the near future (1-3 years).

The goal of this paper is to develop a highly available, real-time data system for public transportation, at a low cost. This system should be easy for public transit agencies to implement, in order to reap the benefits of realtime data with minimal cost.
\section{Prior Work}
OneBusAway \cite{ferris2009onebusaway}, is a set of tools that creates a system for transit traveler information. The system originally included Route Maps, Timetables, a Real-Time Tracker, Service Alerts, and a Trip Planner. Overtime, OneBusAway has been deployed in many places such as Tampa, Florida, Atlanta, Georgia, and New York City, New York. These deployments have created positive results, with 92\% of respondents to a survey reporting that they were more satisfied using public transit result of using OneBusAway. Additionally, users of OneBusAway also reported an increases in public transit trips, allowing for transit agencies to gain more revenue \cite{ferris2010onebusaway}.

Barbeau's research team worked on Pinellas Suncoast Transit Authority's (PSTA) GTFS-RT, located at Pinellas County, Florida. Using this feed, they deployed an instance of OneBusAway for data analysis, discovering three sources of errors. The GTFS-RT producer issues include missing data, along with inconsistent data in relation to the GTFS static data \cite{barbeau2018quality}. Overall, troubleshooting the GTFS-RT feed with the team, agency, and vendor (GTFS-RT producer) was highly challenging, as OneBusAway functions more as a user application rather than a data validator. PSTA staff had to catch errors in real-time through the OneBusAway mobile app, causing a lot of resistance in their debugging.

U.S. Patent No. 20170270790 claimed a ``Vehicle Prediction Module'' system, in which a series of modules are created in order to process inputs of vehicle positions and output ETAs (Estimated Time of Arrival) for transit vehicles using a database of historical arrivals and scheduled stops.

Finally, U.S. Patent No. 10977941 claimed a system where transit terminals, such as payment terminals, use a user device's geolocation as a geolocation for that interacting terminal device \cite{us20170270790a1} \cite{KanMeyer2021}. Then, other users could use that prior geolocation from a user's device via their own devices.
\section{Data Format}
GTFS, General Transit Feed Specification \cite{gtfs_schedule_2024}, is a data format that allows for public transit to be machine-readable to transit application providers.
GTFS can include data such as routes, schedules, and stops. GTFS also has a real-time component \cite{gtfs_realtime_2024}, often shorthanded as GTFS-RT. This component of the specification contains time sensitive data such as vehicle position and trip progress. GTFS-RT data can be created from an AVL (Automatic Vehicle Location) system \cite{oluwatobi1999gps} and a set of GTFS data. In our system, we would have a processing step similar to \cite{ranvcic2008online} to manage the creation of GTFS-RT data.
\section{Sensor Hardware}
The hardware chosen as the sensor is the LILYGO\textregistered T-A7670G R2 development board. The development board features a Espressif\textregistered ESP32-WROVER-E CPU package with 4MB of Flash and 8MB of PSRAM, SIMCOM\textregistered A7670 LTE and GPS modem, and a Quectel\textregistered L76K GPS modem. The GPS modem on the A7670 requires a powered antenna, thus LILYGO\textregistered also attached the Quectel\textregistered L76K GPS modem to provide GPS functionality with a passive antenna. The development board was chosen because of its low price and compatibility with mobile networks in the United States.

Along with the T-A7670G R2 development board, other variants of the development board were tested. The 2 other variants tested include the T-A7080G S3 and the T-A7600G. Differences of the T-A7080G S3 compared to the T-A7670G R2 is the inclusion of a more powerful ESP32-S3 processor and change of modem from A7670 to A7080, which required different handling in firmware for GPS functionality, explained in the next section. The difference of the T-A7600G compared to the T-A7670G R2 is the use of the older A7600 modem, which due to the sunset of 3G networking, is now unable to authenticate with mobile carriers. 
\section{Sensor Firmware}
The primary goals of the sensor firmware were twofold. One, to collect and parse GNSS data to send. Second, the GNSS data is sent via LTE networking to a server for further processing.

\begin{figure}[htbp]
  \centering
  \includesvg[width=0.8\linewidth]{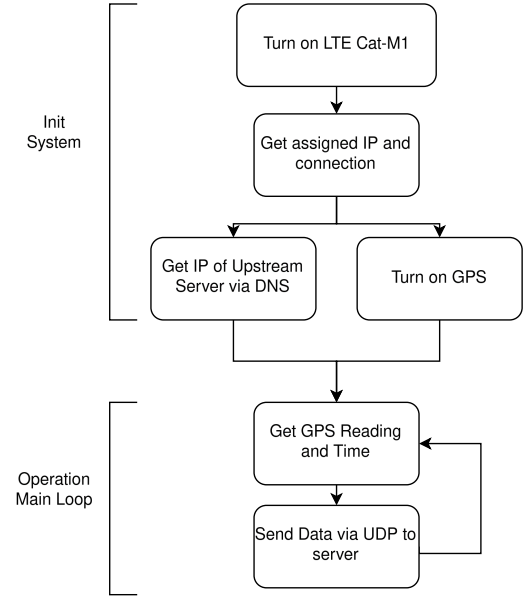}
  \caption{Device with simultaneous GNSS and LTE networking}
  \label{fig:dev}
\end{figure}

The sensor firmware was created for the LILYGO\textregistered T-A7670G R2 with GPS. The A7670G and L76K are logically connected to the ESP32 via serial pins. Each development board is connected to a server cluster via LTE Cat-M1, as they are equipped with an IoTDataWorks 64 kbps Unlimited IoT SIM Card, which uses the T-Mobile US network. The firmware executes the primary steps listed in Figure \ref{fig:dev}.

There is, however, another variant of the firmware designed for devices that cannot receive GNSS and LTE signals simultaneously. The individual steps are similar, but there is no initialization sequence. The init system steps are moved to the main loop, as the device requires both GNSS and LTE modules to be enabled and disabled sequentially to maintain mutual exclusivity.

The first step of each loop (run every 500ms) is to obtain a reading from the L76K GNSS module. The program then extracts GNSS data, using the Arduino Library TinyGPS++ \cite{hart2013tinygpspp}, from the \$GPGGA and \$GPRMC NEMA sentences that the module produces. Another thing collected is the time (as a datetime string) from the mobile network, as GTFS-RT does not need centisecond accuracy that the GNSS module can provide \cite{tumuhairwe2019determining}. This time is then parsed from the string, which is formatted as ``DD/MM/YY,HH:MM:SS±HH'', using another Arduino Library called TimeLib. If the location, bearing, speed, or time is not valid, the loop returns early without going into the second step of the main loop.

\begin{figure}[htbp]
  \centering
  \includesvg[width=\linewidth]{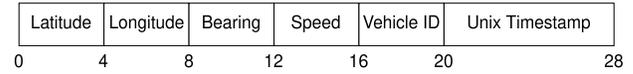}
  \caption{Application Layer Packet Byte Layout}
  \label{fig:packet}
\end{figure}

Next, each loop sends 28 bytes (see \autoref{fig:packet}) to a UDP server, encapsulated by a IPv4/UDP packet. The first eight bytes of the packet are two 32-bit floating-point numbers representing the vehicle's latitude and longitude, followed by its bearing (in degrees) and speed (in kilometer per hours), also as 32-bit floats. Next, there is a vehicle id as a 32-bit unsigned integer. Finally, the timestamp, in Unix epoch format, is the last 8 bytes as a 64-bit unsigned integer. All floating-point numbers in this packet are using the IEEE-754 standard representation.
\vspace{-10pt}
\vspace{-25pt}
\section{Network Architecture}
Our deployment of our demonstration system primarily focused on redundancy and consensus, as it contained the minimum amount of clusters (3) for consensus to be viable.
\begin{figure}[htbp]
  \centering
  \includesvg[width=0.8\linewidth]{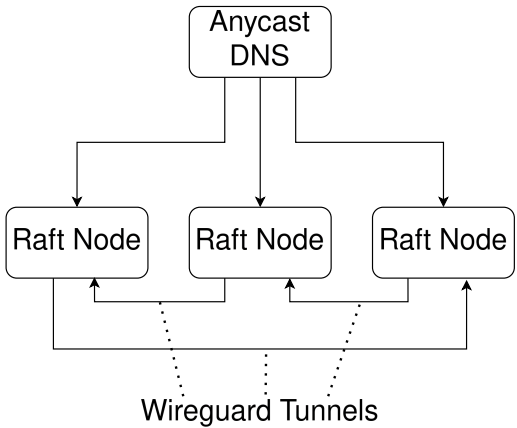}
  \caption{Cluster Network Topology}
  \label{fig:arch}
\end{figure}
We separated our nodes physically into separate networks, each with unique IPv4 addresses, so that we can use Anycast DNS in order to route requests users send to us to the nearest node. Each node was also connected via a WireGuard Mesh VLAN because our chosen consensus algorithm, Raft \cite{ongaro2015raft}, is not Byzantine Fault Tolerant \cite{copeland2016tangaroa}. To maintain the security of our service, we deployed Raft within a private network where its service is not exposed publicly.

\section{Server Software}
\textit{Given the position of a transit vehicle, its ID, the time, the route ID, and the GTFS Schedule data, can we determine a proper VehiclePosition for the vehicle?}
\subsection{Preprocessing}
The server has three primary data structures (Fig.  \ref{fig:server}) that need to be filled with preprocessed data before the primary threads of the server can run. First, the GTFS-RT message stored in global memory is initialized with the current timestamp and header data. Next, the K/V map that is used for in-memory queries of vehicle and route attachment data is initialized with preset data from a file. Finally, a GTFS file is preprocessed into an interval tree, where each of its trips are stored in nodes. This allows the program to do trip queries in $O(\log n)$, where each trip is in a given interval.
\subsection{Architecture}
\begin{figure}[h]
  \centering
  \includesvg[width=0.8\linewidth]{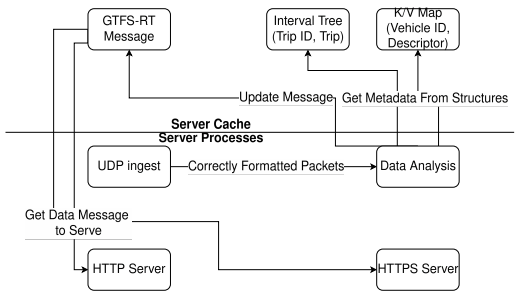}
  \caption{Internal Server Architecture}
  \label{fig:server}
\end{figure}

In Figure \ref{fig:server}, the architecture of the server-side software of this system is shown. The figure shows how the server handles the data flow and processing pipeline of converting the packet format in Figure \ref{fig:packet} received from GPS sensors, into GTFS-RT messages. Since GTFS-RT is a standard, we will use it as a data format to provide live information regarding our transit system. The architecture revolves around a central server cache and includes four primary threads that do the following: UDP ingestion, data analysis, and HTTP/HTTPS service. All of these threads work in parallel to receive, process, and serve real-time transit data efficiently.

\subsection{UDP Ingestion}
The process begins with a UDP ingest thread, which listens for incoming UDP packets from external sources, likely vehicle tracking systems or transit agencies. These packets contain raw vehicle position data or other GTFS-RT message components. Once received, these packets are analyzed in order to determine their validity before being sent to the data processing thread. We do not want our feed to contain invalid data, which is why this step is required.
\subsection{Data Analysis}
The validated packets are sent to to the Data Analysis thread. It performs operations on the correctly formatted packets, by extracting the data and correlating it with previously set data, such as a route ID for their vehicle. For this correlation, the system queries a Key/Value Map, which contains mappings between vehicle IDs and their descriptors, which is metadata about the vehicle including their id, label, license plate, and wheelchair accessibility. Then, this data is constructed into a valid VehiclePosition GTFS-RT message using Algorithm \ref{algo:gps}.

Algorithm~\ref{algo:gps} generates a GTFS-RT VehiclePosition message from a GPS reading, vehicle ID, and timestamp. It cross-references a shared vehicle-to-route mapping and a time-indexed trip schedule to determine the most probable trip the vehicle is operating on. When multiple trips are possible, it selects the most likely one using spatial indexing via a KDTree \cite{wong2025algorithmicanalysisgtfsrtvehicle}. If the vehicle is within 20 meters of a stop, it includes that stop in the message and assigns a \texttt{StoppedAt} status if the vehicle's speed is zero.

\begin{algorithm}[htbp]
\caption{GPS Data to GTFS-RT Vehicle Position}
\label{algo:gps}
\SetAlgoLined
\DontPrintSemicolon
\KwIn{Position: (Latitude, Longitude, Bearing, Odometer, Speed), Vehicle ID, Timestamp}
\KwOut{VehiclePosition}

Acquire lock on Shared Vehicle Map\;
Lookup \texttt{(vehicle, route\_id)} using Vehicle ID\;
Release lock on Shared Vehicle Map\;

\If{route\_id is None}{
    \Return VehiclePosition with minimal fields filled (no trip info)\;
}

Compute \texttt{seconds\_since\_midnight} from timestamp\;

Acquire lock on Shared Trip Tree\;
Get all trips overlapping \texttt{seconds\_since\_midnight}\;
Release lock on Shared Trip Tree\;

\ForEach{trip in trips}{
    Find closest stop before and after current time\;
    Collect stops between these two points\;
    Store as tuple (trip\_id, stop\_times)\;
}

\If{all trip\_ids are equal}{
    Set \texttt{trip\_id} to that common value\;
}
\Else{
    For each trip candidate:\;
    \Indp
    Build 2D KDTree with stop coordinates\;
    Find nearest stop to position;\;
    Store closest trip\_id\;
    \Indm
    Pick trip with nearest stop as final \texttt{trip\_id}\;
}

Find full trip object matching \texttt{trip\_id}\;

Build KDTree of all stops in selected trip\;
Find nearest stop to current position\;

\Return VehiclePosition with trip, stop\_id, current\_stop\_sequence, and vehicle info\;
\end{algorithm}

The VehiclePosition message then is sent to update the GTFS-RT Message, which is the final result for all real-time data. This GTFS-RT Message is kept up to date with the most recent information and is what the HTTP/HTTPS servers return when they are sent requests.

\subsection{HTTP/HTTPS Service}
A client can request GTFS-RT data, via HTTP or HTTPS. This will cause the HTTP Server or HTTPS Server retrieve the GTFS-RT Message from the server cache and then return it to the client.
\section{Discussion and Conclusion}

This work presents the design and implementation of a GTFS-RT positions system, from sensor hardware and embedded firmware, along with a server-side data for distributed feed delivery. Our system demonstrates that it is feasible to build a low-cost, highly available solution for real-time public transit data, tailored for small and mid-sized agencies that may lack significant technical or financial resources. By building a modular and open architecture, we lower the barrier to entry for transit operators and simplify maintenance and deployment.

Compared to prior work like OneBusAway and vendor-provided systems, our approach focuses on physical system-level control and transparency, allowing agencies to generate and troubleshoot their own GTFS-RT feeds with confidence. Our design choices, such as choosing open standards, using minimal and energy-efficient hardware, and distributing load across cloud regions, directly support our goals of cost-effectiveness, reliability, and long-term scalability. While patent literature often focuses on proprietary prediction systems or sensor-device interactions, our contribution centers around public, standards-compliant data feeds that any user or app can access.

Finally, there are still areas for future work. For example, GTFS-RT has data fields for real-time ETA predictions, which require using machine learning models.
Furthermore, we could also work on supporting other types of GTFS-RT feeds, such as Alerts and TripUpdates. Nevertheless, our system establishes a functional, end-to-end foundation that aligns with the growing expectations for real-time data in public transportation. As public pressure and regulatory guidance increase in favor of open data, we believe systems like ours will play a crucial role in expanding GTFS-RT adoption across smaller transit operators in California and beyond.
\begin{acks}
Thank you Hacker Initiative for sponsoring this work through Lavender Computing Collective, a 501(c)(3) non-profit corporation.
\end{acks}

\bibliographystyle{ACM-Reference-Format}
\bibliography{paper}

\end{document}